\documentclass[american,english,journal=jctcce,manuscript=article]{achemso}

\usepackage{xcolor}
\usepackage{amsmath}

\newcommand{\tf}{\textcolor{blue} }

\title{Automatic learning of hydrogen-bond fixes in an AMBER RNA force field}

\author{Thorben Fr\"ohlking}
\affiliation{Scuola Internazionale Superiore di Studi Avanzati, via Bonomea 265, 34136 Italy}%
\alsoaffiliation{Equal contribution}
\author{Vojt{\v e}ch Ml{\'y}nsk{\'y}}
\affiliation{Institute of Biophysics of the Czech Academy of Sciences, Kralovopolska 135, 612 65 Brno, Czech Republic}
\alsoaffiliation{Equal contribution}
\author{Michal Jane{\v c}ek}
\affiliation{Department of Physical Chemistry, Faculty of Science, Palacky University, tr. 17 listopadu 12, 771 46, Olomouc, Czech Republic}
\author{Petra K{\"u}hrov{\'a}}
\affiliation{Regional Centre of Advanced Technologies and Materials, Czech Advanced
Technology and Research Institute (CATRIN), Palacky University Olomouc,
Slechtitelu 27, 779 00 Olomouc, Czech Republic}
\author{Miroslav Krepl}
\affiliation{Institute of Biophysics of the Czech Academy of Sciences, Kralovopolska 135, 612 65 Brno, Czech Republic}
\alsoaffiliation{Regional Centre of Advanced Technologies and Materials, Czech Advanced
Technology and Research Institute (CATRIN), Palacky University Olomouc,
Slechtitelu 27, 779 00 Olomouc, Czech Republic}
\author{Pavel Ban{\'a}{\v s}}
\affiliation{Regional Centre of Advanced Technologies and Materials, Czech Advanced
Technology and Research Institute (CATRIN), Palacky University Olomouc,
Slechtitelu 27, 779 00 Olomouc, Czech Republic}
\author{Ji{\v r}{\' \i} {\v S}poner}
\affiliation{Institute of Biophysics of the Czech Academy of Sciences, Kralovopolska 135, 612 65 Brno, Czech Republic}
\alsoaffiliation{Regional Centre of Advanced Technologies and Materials, Czech Advanced
Technology and Research Institute (CATRIN), Palacky University Olomouc,
Slechtitelu 27, 779 00 Olomouc, Czech Republic}
\author{Giovanni Bussi}
 \email{bussi@sissa.it}
\affiliation{Scuola Internazionale Superiore di Studi Avanzati, via Bonomea 265, 34136 Italy}%

\begin{document}
\begin{abstract}
The capability of current force fields to reproduce RNA structural dynamics  is limited.
Several methods have been developed to take advantage of experimental data in order to enforce agreement with experiments. We herein extend an existing framework,
which allows arbitrarily chosen force-field correction terms to be fitted by quantification of the discrepancy between observables back-calculated from simulation and corresponding experiments.
We apply a robust regularization protocol to avoid overfitting, and additionally introduce and compare a number of different regularization strategies, namely L1-, L2-, Kish Size-, Relative Kish Size- and Relative Entropy-penalties.
The training set includes a GACC tetramer as well as more challenging systems, namely gcGAGAgc and gcUUCGgc RNA tetraloops.
Specific intramolecular hydrogen bonds in the AMBER RNA force field are corrected with automatically determined parameters that we call gHBfix$_{opt}$.
A validation involving
a separate simulation of a system present in the training set (gcUUCGgc) and new systems not seen during training (CAAU and UUUU tetramers) displays improvements regarding native population of the tetraloop as well as good agreement with NMR-experiments for tetramers when using the new parameters.
Then we simulate folded RNAs (a kink-turn and L1 stalk rRNA) including hydrogen bond types not sufficiently present in the training set. This allows a final modification of the parameter set which 
is named gHBfix21 and is suggested to be applicable to a wider range of RNA systems.

\end{abstract}

\maketitle

\section{Introduction}

\begin{figure*}
  \includegraphics[width=\columnwidth]{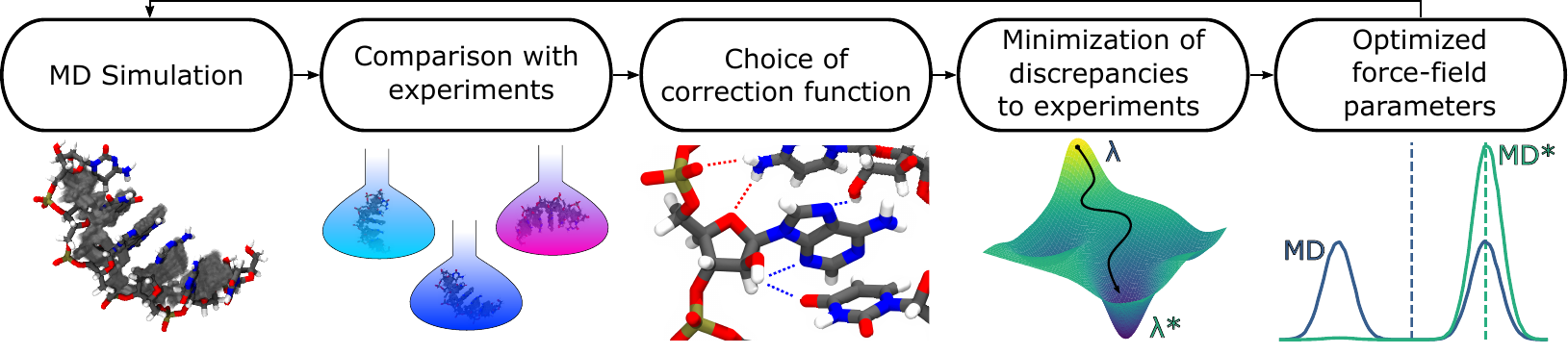}
  \caption{Schematic visualization of the workflow for automatic force-field refinement \cite{frohlking2020toward}.
After performing MD simulations on training systems for which experimental data are available,
experimental quantities are back-calculated and compared with actual experimental data points.
One then chooses a basis set for the correction function.
The gHBfix corrections \cite{kuhrova2019GHBFIX} are a natural choice to compensate for the possibly incorrect relative stability
of hydrogen bonds in the AMBER force field.
A numerical minimization is then performed so as to maximize the agreement between
simulation and experiment, based on reweighting the simulated trajectories.
Ideally, the resulting force field parameters enable new simulations to generate
structural ensembles in better agreement with experiment also for systems not included in the training set.
If necessary, a new minimization can be performed using a combination of the original and new trajectories,
in an iterative fashion.
\label{fig_overview}}
\end{figure*}

\begin{figure}
\includegraphics[width=0.45\textwidth]{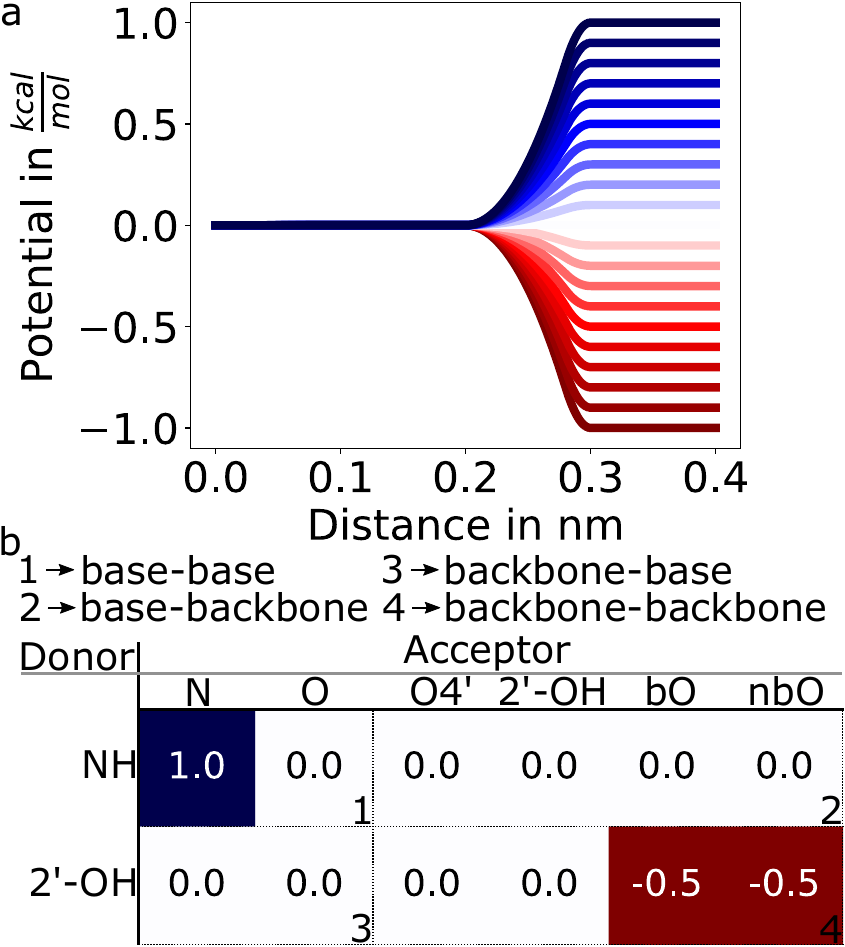}
\caption{
(Panel a) Functional form for the gHBfix-correction potential \cite{kuhrova2019GHBFIX}, displayed as a function of the distance between a hydrogen and the corresponding acceptor.
The color scale indicates corrections that could either support (blue) or disfavor (red) a hydrogen bond type.
(Panel b) In the present work six possible acceptors and two possible donors are systematically considered, leading to a total of twelve trainable parameters. The numbers show the initial set of parameters ($\eta$ parameters proposed in Ref. \cite{kuhrova2019GHBFIX}, also referred to as gHBfix19 parameters), here expressed as $k_BT\cdot \lambda$ in kcal/mol.
6 possible acceptors and 2 possible donors are considered in this work, thus leading to a total of 12 trainable parameters.
Parameters are colored in blue or red according to the same scale used in panel a.
\label{fig_parameters}
}
\end{figure}

As viral pandemics are approached with RNA vaccines \cite{teijaro2021covid} and
RNA is becoming an increasingly relevant target in therapeutics \cite{matsui2017non},
accurate methods for predicting and designing structures and dynamics of nucleic acids are needed to accelerate progress in these fields. Molecular dynamics (MD) simulations in principles allow RNA dynamics
to be modeled by computing interactions using empirical force fields and directly solving the equations of motion.
However, the capability of MD simulations to predict RNA dynamics is limited both by sampling issues and by force-field accuracy
\cite{sponer2018rna}.
Depending on the size of the system and on the complexity of the investigated conformational transitions, enhanced sampling
techniques \cite{bernardi2015enhanced,mlynsky2018exploring} can help decrease the time-scale issue significantly. However, especially when long simulation time scales or enhanced-sampling methods are employed, the accuracy of the underlying force fields can become a critical issue and can lead to
structural ensembles that do not agree with experiment for disordered oligomers \cite{condon2015stacking,bergonzo2015highly} or for difficult structural motifs \cite{bottaro2016free,mrazikova2020uucg}.
A number of possible approaches can be used to take advantage of available experimental data in order to enforce agreement between experiments and simulation data
\cite{norgaard2008experimental,li2011iterative,cesari2016combining,cesari2019fitting,kofinger2021empirical}
(see also Refs.~\cite{orioli2019learn,frohlking2020toward} for recent reviews and Fig.~\ref{fig_overview} for a schematic).
Critical and partly related issues in the application of these methods are (a) avoiding overfitting, which can be moderated by using
properly tuned regularization terms\cite{cesari2019fitting,frohlking2020toward},
and (b) explicitly modeling experimental errors, which can be naturally done in Bayesian formulations
\cite{kofinger2021empirical}.
Both approaches require the degree of confidence one has in experiments and simulations
to be tuned.
These approaches are expected to generate transferable force fields, and should not be confused with
non-transferable ensemble refinements that aim at minimal ensemble corrections without requiring transferability
of the resulting force-field form (see, e.g., Refs.~\cite{pitera2012use,white2014efficient,hummer2015bayesian,bonomi2016metainference,cesari2018using}). In particular, 
approaches for transferable force-field refinement are dependent on the functional form of the
correction terms to be fixed a priori using chemical intuition.
For atomistic MD simulations, these corrections could for instance act on dihedral angle potentials
\cite{li2011iterative,cesari2019fitting}. This is a natural choice, since dihedral angles
are usually fitted as a last step, are expected to compensate for all the errors accumulated in other
force-field terms, and are naturally connected to the population of different rotamers \cite{richardson2008rna}.
However, recent works suggested that an imbalance in the relative strength of
solute-solute hydrogen bonds might be a key problem of current RNA force fields
so that fixing these terms might be more effective than
acting on dihedral angles \cite{kuhrova2019GHBFIX,mlynsky2020fine,mrazikova2020uucg}.
In these works, a limited number of hydrogen bond types were corrected using a so-called
generalized hydrogen-bond fix (gHBfix, see Fig.~\ref{fig_parameters}), with promising results.
This approach allows for minimal corrections that are less likely to present side effects when compared to
more extensive reparametrizations of non-bonded interactions \cite{tan2018rna} as shown in Ref.~\citep{kuhrova2019GHBFIX}.
Correction factors for the gHBfix force field,
leading to either supporting or disfavoring specific hydrogen bond types, were chosen by trial and error, using a protocol
that might be difficult to generalize \cite{kuhrova2019GHBFIX}.

In this paper we expand on this idea and show that it is possible to train in an automatic fashion
the correction factors associated to hydrogen-bond stabilization in the gHBfix model so as to stabilize
the native structure of the difficult \cite{mrazikova2020uucg} UUCG tetraloop structural motif.
To avoid overfitting on the UUCG tetraloop, a tetraloop representative of a different class and a flexible tetramer are
included in the training set.
We use an approach heavily based on that reported in Ref.~\citep{cesari2019fitting}.
As an important extension, here we introduce and compare (a) different forms of the regularization term
and (b) different protocols that can be used to perform cross-validation.
Training of the 12 parameters of the gHBfix force field is done
and demonstrates that this approach can lead to
transferable and interpretable force field corrections  that match experimental data
on a range of systems.
A critical assessment of the 
side effects of the optimized corrections is made. Further tests on carefully chosen folded RNAs 
allow us to design a final set of parameters
(gHBfix21) that is transferable on a wide range of RNA structural motifs.
This is an upgrade of the set suggested earlier \cite{kuhrova2019GHBFIX},
which is also known as gHBfix19 \cite{mrazikova2020uucg}.
Similarly to the preceding gHBfix19 variant, gHBfix21 should be coupled with the OL3 force field \cite{Cornell996,Wang2000,Perez2007,Zgarbova2011}
with modified phosphate parameters \cite{Steinbrecher2012,Mlynsky2015} and OPC water model \cite{IzadiOPC2014} as used here.

\section{Methods}

\subsection{Simulation protocols}

We performed simulations of several RNA systems, namely (i) GACC, CAAU, and UUUU tetranucleotides, (ii) gcGAGAgc and gcUUCGgc 8-mer tetraloops, (iii) ggcacUUCGgugcc 14-mer tetraloop (PDB ID 2KOC \cite{nozinovic2010high}),
(iv) gcaccguugg (PDB ID 1QC0\cite{klosterman1999crystal}) and uuauauauauauaa (PDB ID 1RNA \cite{dock1989crystallographic}) RNA duplexes, (v) Kink-turn (Kt-7, PDB ID 1S72\cite{klein2004roles}, \tf{19 nucleotides}) and L1 stalk rRNA (PDB ID 3U4M\cite{Tishchenko2012}, \tf{80 nucleotides}) motifs.
Starting structures of tetranucleotides and 8-mer tetraloops (in unfolded states) were prepared using Nucleic Acid Builder of AmberTools14 \cite{case2005amber} as one strand of an A-form duplex. The topology and coordinates of simulated systems were prepared using the tLEaP module of AMBER16 program package.\cite{amber16}
Several trajectories for analysis were taken from our previous works (see SI Table 1 %
 for full list of systems and simulations). All systems were solvated using a rectangular box of OPC \cite{IzadiOPC2014} water molecules with a minimum distance between box walls and solute of 12 \AA. We used the standard OL3 RNA ff \cite{Cornell996,Wang2000,Perez2007,Zgarbova2011}  with the vdW modification of phosphate oxygens developed in Ref.~\citep{Steinbrecher2012} where the affected dihedrals were adjusted as described elsewhere \cite{Mlynsky2015}. AMBER library file of this ff version can be found in Supporting Information of Ref. \citep{kuhrova2016ghbfixfunc}. Standard MD simulations were run at $\approx$0.15 M KCl using the Joung-Cheatham ion parameters \cite{CheathamIons2008} (K$^+$: r = 1.705 \AA, $\epsilon$ = 0.1937 kcal/mol, Cl$^-$: r = 2.513 \AA, $\epsilon$ = 0.0356 kcal/mol). Enhanced sampling simulations of tetranucleotides and tetraloops were run at $\approx$0.15 M and $\approx$1.0 M KCl salt excess, respectively. We used the hydrogen mass repartitioning scheme \cite{Hopkins2015} allowing a 4-fs integration time step (see Supporting Information of ref. \citep{kuhrova2019GHBFIX} for other details about the simulation protocol). Hydrogen bonds were tuned by various versions of the gHBfix potential \cite{kuhrova2019GHBFIX} (Table 2 in the reference and SI Table 1 of this study). Standard MD simulations were run in AMBER18 \cite{amber18} whereas both AMBER18 and GROMACS2018 \cite{abraham2015gromacs} were used for enhanced sampling simulations. PARMED \cite{Shirts2017} was used to convert AMBER topologies and coordinates into GROMACS inputs.
Two different enhanced sampling schemes were employed, i.e., standard replica exchange solute tempering (REST2)\cite{Wang2011} protocol and well-tempered Metadynamics
\cite{laio2002escaping,barducci2008well,bussi2020using} (MetaD)
in combination with REST2 method (ST-MetaD) \cite{Camilloni2008,mlynsky2021towards}. REST2 simulations were performed at 298 K (the reference replica) with 8 and 16 replicas for tetranucleotides and UUCG 8-mer tetraloop, respectively.
Details about settings can be found elsewhere \cite{kuhrova2019GHBFIX}. The scaling factor (lambda) values ranged from 1 to 0.601700871 and from 1.0454 to 0.59984 for 8 and 16 replicas, respectively. Those values were chosen to maintain an exchange rate above 20\%. The effective solute temperature ranged either from 298 K (8 replicas) or 285 K (16 replicas) to ~500 K. REST2 simulations were performed with AMBER GPU MD simulation engine (pmemd.cuda) \cite{Mermelstein2018}.
ST-MetaD simulations of both GAGA and UUCG 8-mer tetraloops were performed with 12 replicas starting from unfolded single strands and were simulated in the effective temperature range of 298-497 K for 5 $\mu$s per replica. The average acceptance rate was ~30\% for both tetraloops. The eRMSD metric \cite{bottaro2014role} was used as a biased collective variable \cite{bottaro2016free}. We used eRMSD with an augmented cutoff (set at 3.2) for biasing, which was shown to allow forces to drive the system towards and away from the native state even when nucleobases are far from each other \cite{bottaro2016free}.
In a separate manuscript, we show that using ST-MetaD with MetaD on eRMSD greatly improved the performance of pure ST for RNA tetraloops \cite{mlynsky2021towards}.
Similar conclusions were drawn in Ref. \cite{zerze2021}, where parallel tempering-MetaD \cite{bonomi2010enhanced,bussi2006free} with MetaD
on the number of native contacts \cite{best2013native}, a variable highly correlated with eRMSD, was suggested to be significantly more efficient than pure parallel tempering
for a GNRA tetraloop.
ST-MetaD simulations were carried out using GPU-capable version of GROMACS2018 \cite{abraham2015gromacs} in combination with PLUMED 2.5 \cite{Tribello2014,bussi2014hamiltonian} (see Section \ref{ghbfixImpl} for more details about implementation of the gHBfix function within PLUMED code and Table SI 1 in Supporting Information for full list of standard as well as enhanced sampling simulations).
In theory, all replicas could be combined using a suitable reweighting procedure. However,
to keep the datasets smaller, we here decided to only analyze the reference replica of each replica-exchange simulation.
Simulations for the same system performed with different force fields were combined with binless
weighted-histogram analysis \cite{Souaille2001,Shirts2008,Tan2012} so as to
maximize the statistical efficiency of the reweighting procedure.

\subsection{Experiment-based force-field fitting}

We briefly review the formalism behind experiment-based force-field fitting. We here used the procedure discussed in Ref.~\cite{cesari2019fitting}. Considering $P_0(x)$ as the equilibrium probability distribution of observing a conformation $x$ with the original force field,
the refined force field will include a correction in the form $f(x,\{\lambda\})$, where
$\{\lambda\}$ is a set of $N$ parameters, leading to an equilibrium distribution
$
P(x,\{\lambda\}) \propto P_0(x) e^{-f(x,\{\lambda\})}
$.
We here assume that the correction $f(x,\{\lambda\})$ is a linear combination of $N$ correction functions:
$
f(x,\{\lambda\}) =
\sum_{j=1}^N \lambda_j f_j(x)
$.
The modified distribution is then used to estimate the expectation value of $M$ experimental observables, defined 
through forward models $O_i(x)$ that connect the atomic coordinates of conformation $x$ with the experiment.
Forward models might correspond for instance to Karplus equations \cite{karplus1963vicinal} or to
indicator functions equal to 1 if $x$ is a folded conformation and to 0 otherwise.
Their expectation values are computed as
$\langle O_i \rangle (\{\lambda\}) = \sum_x O_i(x) P(x,\{\lambda\})$.
The cost function, to be minimized in the fitting procedure, can be written as an average of
squared discrepancies between these expectation values and the corresponding experimental observations:
\begin{equation}
\chi^2(\{\lambda\}) = \frac{1}{M}\sum_{i=1}^M \left(\frac{
\langle O_i \rangle (\{\lambda\}) - O_i^{exp}
}{
\sigma_i
}
\right)^2
\label{eq:cost}
\end{equation}
Here $\sigma_i$ is an estimate of the experimental error associated to the $i$-th datapoint.

In this work, the functions $f$ are defined following the gHBfix potential function as formulated in ref.~\cite{kuhrova2019GHBFIX}.
In our implementation, the parameters $\{\lambda\}$ are unitless.
However, when reporting them in figures and tables, we convert them to kcal/mol units for clarity,
by multiplying them by $k_BT$ where $k_B$ is the Boltzmann constant and $T$ is the simulation temperature.
Each of the fitted parameters thus report on how much a given hydrogen bond type is supported (positive)
or disfavored (negative).

\subsection{Back-calculation of experimental observables}

For the two tetraloops, we identified the frames corresponding to native structures
using the same procedure used in Ref.~\cite{kuhrova2019GHBFIX}, namely
considering frames with
eRMSD \cite{bottaro2014role} from native smaller than 0.7 \cite{bottaro2019barnaba} and all native hydrogen bonds
(Watson-Crick hydrogen bonds in the stem and signature hydrogen bonds in the loop,
see Figure 1 in Ref.~\citep{kuhrova2019GHBFIX}).
Calling $p_i$ the population of the native state for system $i$, a contribution to
the cost function is computed as:
\begin{equation}
\chi_{i}^2(\{\lambda\})=\left(\log \text{min}\left(\frac{p_i(\{\lambda\})}{0.5},1\right)\right)^2\,.
\label{eq:cost-nativepop}
\end{equation}
In this manner, a penalty is added whenever the native population is lower than 50\%.

For the tetramers we computed the agreement with previously published NMR data\cite{Condon2015,Turner2020,bottaro2018conformational}
using the same procedure as in Ref.~\cite{kuhrova2019GHBFIX}.%

\subsection{Fitting on multiple systems}

The procedure above can be straightforwardly generalized to multiple systems.
In practice, separate
error functions are computed for each system and their linear combination is taken.
Explicitly, if $\chi^2_i(\{\lambda\})$ is the error function for the $i$-th system, computed using Eq.~\ref{eq:cost},
the total cost function over $S$ systems can be defined as
\begin{equation}
\chi^2(\{\lambda\})=\sum_{i=1}^S \omega_i \chi^2_i(\{\lambda\})
\label{eq:cost-multi}
\end{equation}

The prefactor associated to each system in this linear combination ($\omega_i$) allows the weight of each system in the fitting procedure to be tuned.
Each of the 3 systems considered in this study is assigned the same weight $\omega_i=1$,
so that they equally contribute to the overall error.
Note that these parameters have to be chosen arbitrarily and might have significant impact on the combined $\chi^2$.

\subsection{Regularization terms}

The cost function in Eq.~\ref{eq:cost-multi} can be augmented with a regularization term so as to decrease the degree of overfitting:
\begin{equation}
\tilde{\chi}^2(\{\lambda\})=
\chi^2(\{\lambda\}) + \alpha R(\{\lambda\})
\end{equation}
Here $R$ is a function that takes into account how much the force field has been fitted and thus
typically grows as the refined force field departs from the reference one.
$\alpha$ is a regularization hyperparameter that can be tuned using a cross-validation procedure.
We here compare a number of different functional forms for the regularization function $R(\{\lambda\})$.
The most common type of regularization is L2 regularization, where the function $R$ is defined as
\begin{equation}
R(\{\lambda\})
= L_2(\{\lambda\})
=
\sum_{i=1}^N (\lambda_i-\lambda_i^0)^2
\end{equation}
where $\{\lambda^0\}$ are the parameters suggested in the original work \cite{kuhrova2019GHBFIX},
shown in Fig.~\ref{fig_parameters}.
This type of regularization corresponds to setting a Gaussian prior on the parameters $\{\lambda\}$.
Indeed, the logarithm of a Gaussian function of the $\{\lambda\}$'s is proportional to a quadratic
function of the $\{\lambda\}$'s.
Similarly, a Laplace pior would result in a L1 regularization:
\begin{equation}
R(\{\lambda\})
= L_1(\{\lambda\})
=
\sum_{i=1}^N |\lambda_i-\lambda_i^0|
\end{equation}
L1 regularization leads to more sparse corrections than the L2 regularization, meaning it also offers the potential to identify the most important parameters. In addition to comparing L1 and L2 regularization functions, we also tested a function that depends on the statistical significance of the generated ensemble, namely the inverse of the Kish sample size of the reweighted trajectory \cite{GrayKish1969,rangan2018determination}:
\begin{equation}
R(\{\lambda\})
= \frac{1}{K(\{\lambda\})}
= \sum_t w_t^2
\end{equation}
Here $w_t$ depends on $\{\lambda\}$ and represent the reweighting factor for the $t$-th frame, namely
$w_t=\frac{w_{t0}e^{-f(x(t),\{\lambda\})}}{\sum_{t'} w_{t'0}e^{-f(x(t'),\{\lambda\})}}$,
where $w_{t0}$ is the weight associated to the original force field, included here to take into account
that simulations might have included a bias potential.
We notice that a term depending on the Kish size, though different from this one, was also employed in a recent
work \cite{kofinger2021empirical}.

We then considered regularization terms that take into account the discrepancy between the prior distribution
$P_0(x)$ and the posterior one $P(x)$. We tested the inverse of the relative Kish size, defined as
\begin{equation}
R(\{\lambda\})
= \frac{1}{K_{rel}(\{\lambda\})}
= \frac{1}{N_f}\sum_t \frac{w_t^2}{w_{t0}}
\end{equation}
We also considered the exponential of the negative relative entropy, defined as
\begin{equation}
R(\{\lambda\})
= e^{-S_{rel}(\{\lambda\})}
= e^{\sum_t w_t \log \frac{w_t}{w_{t0}}}
\end{equation}
Although these last two forms are different, they are both expected to grow as the distribution associated to the original force field
and that associated to the refined force field depart from each other.

Since the last three regularization terms depend on the analyzed trajectories, they should be combined 
so as to take into account of how each system is affected by the corrections.
We decided to combine them with a LogSumExp (LSE) function,
\begin{equation}
LSE(\{\lambda\})
= \log{\sum_{i=1}^S e^{R_i(\{\lambda\})}}
\end{equation}
that effectively picks the largest regularization across all systems. This makes sure that all
systems have a sufficient Kish size or a sufficient similarity with the initial ensembles.

The five regularization terms discussed above were modulated by a hyperparameter $\alpha$.
Although in principle they could be combined, we only tested one regularization type at a time.
In addition, we added boundaries for the parameter minimizations relative to the
reference parameters (Fig.~\ref{fig_parameters}).
These boundaries can be interpreted as a L-infinite regularization term
($R=\sum_{i=1}^N\left(\frac{\lambda_i-\lambda_i^0}{\lambda_{max}}\right)^{\infty}$ with $k_BT\lambda_{max}=1$kcal/mol)
used on top of one of the five regularization strategies discussed above.
In practice, these boundaries avoid divergence in parameters that could become arbitrarily positive or negative.

\subsection{Cross-validation strategies}

The hyperparameter that tunes the regularization terms discussed in the previous section is chosen so as to
maximize the performance in cross-validation. In particular, force field corrections are fitted on a fraction of the
available dataset and tested on the left out part of the dataset. We performed three types of cross-validation:
\begin{enumerate}
\item Cross-validation on trajectory segments. We split each trajectory in 5 segments.
The number of segments is here chosen arbitrarily, and only the ground replica is used, that might contain spurious correlations due to replica exchanges.
In principle, one might apply the splitting on continuous (demuxed) trajectories to minimize correlations, and optimize the number of blocks as it is usually done in
block analysis \cite{flyvbjerg1989error}.
Then, we minimize the cost function using trajectories where one of the segments was removed, and finally validate the parameters by recomputing the cost function
using only the left out segments.
\item Cross-validation on observables.
We split the dataset into the 7 observables (GACC: NOEs, uNOEs and scalar-couplings grouped according to the backbone angles $\gamma$ (backbone1), $\beta$ or $\epsilon$ (backbone2) and sugar torsional angles (sugar), GAGA and UUCG: native population) then minimize the cost function in which the contribution of one observable is ignored and afterwards validate against this left out observable.
\item Cross-validation on systems. We minimize the cost function
only including two of the three training systems and then validate the parameters by recomputing the cost function
using only the left out system.
\end{enumerate}

In all cases, the cross-validation is repeated by rotating the left-out portion of the data.
The first cross-validation strategy allows to check if the parameters would be transferable to a new trajectory simulated
for the same set of systems. The other two strategies instead check the transferability to different types of observables
or to different systems.

\subsection{Implementation}
\label{ghbfixImpl}

To allow the gHBfix corrections to be used in generic MD codes that might not support the required functional form,
we added an implementation within the PLUMED plugin \cite{Tribello2014} that is compatible with a large number of MD packages.
Specifically, a collective variable has been added that allows the user to provide two groups of atoms
and then automatically compute switching functions ranging from -1 (small distance) to 0 (large distance) with
a smooth interpolation in the middle.
The decision to set this correction to zero for atoms at large distance was taken to make this function compatible
to other switching functions implemented in PLUMED and enabling its optimization via neighbor lists.
This definition is identical to the one used in the original gHBfix version~\cite{kuhrova2016ghbfixfunc}
except for an additive constant. Multiplicative prefactors for the switching functions can be chosen
based on the atom types.
This collective variable can be used to analyze hydrogen bond interactions \emph{a posteriori} or to generate bias potentials to correct
a simulation on-the-fly.%

We here also updated the code 'gHBfix\_GenerateInput.cpp' originally published in Ref. \cite{kuhrova2019GHBFIX} (\url{https://github.com/bussilab/ghbfix-training}), which is printing desired output with the newly implemented gHBfix function for PLUMED code with both required external files (typesTable.dat, scalingParameters.dat). PLUMED input files used in this code are
available on PLUMED-NEST (\url{https://www.plumed-nest.org}), the public repository of the PLUMED consortium \cite{bonomi2019promoting}, as plumID:21.051.

\section{Results}

\begin{figure}
\includegraphics[width=0.45\textwidth]{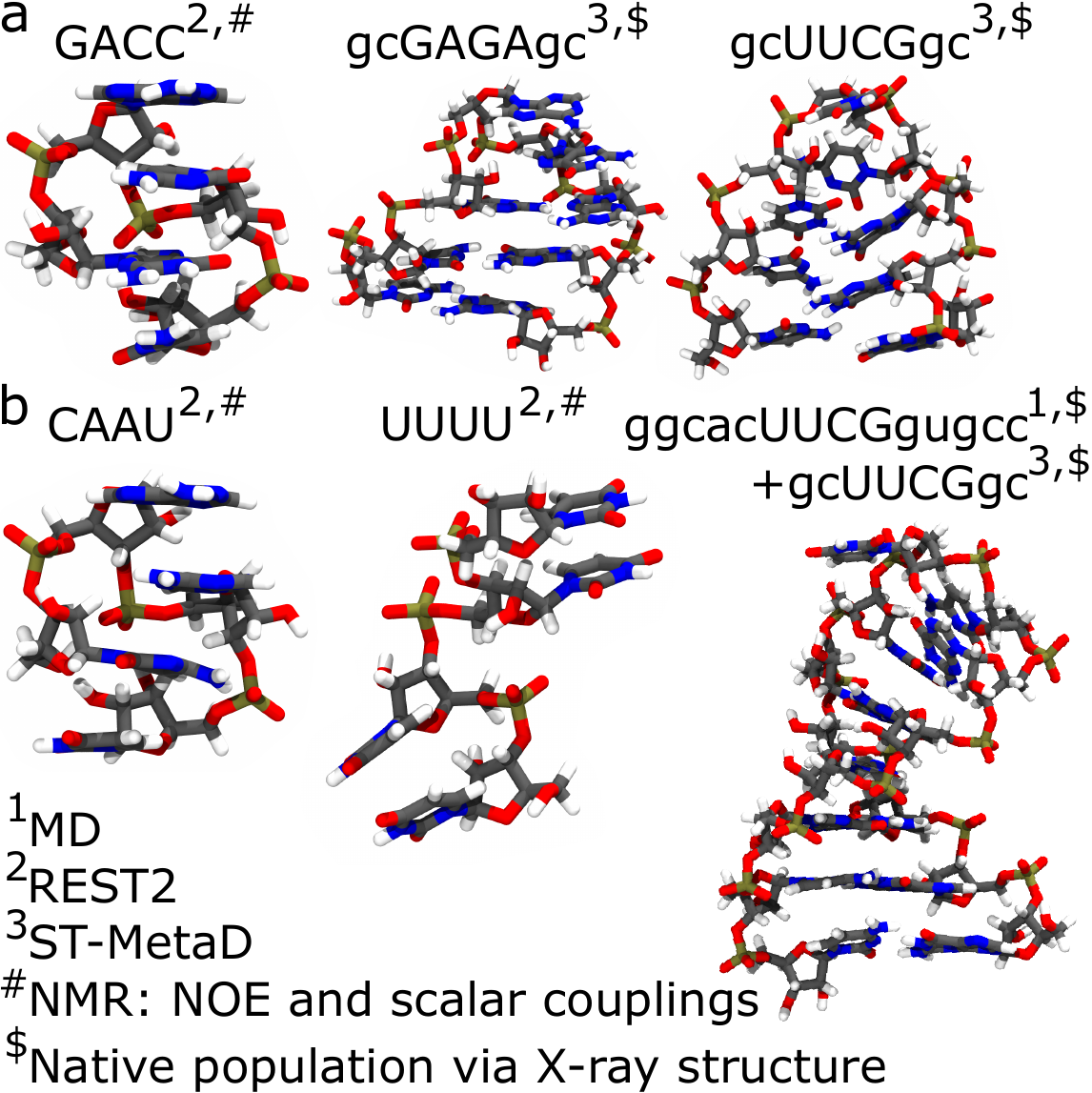}
\caption{
Systems used in training (panel a). For these three systems we performed extensive enhanced sampling simulations.
Training was done using NMR data (for GACC) or stability of native structure (for gcGAGAgc and gcUUCGgc).
Systems used in validation (panel b). Quantitative validation was done using NMR data (for CAAU and UUUU)
or stability of native structure (for gcUUCGgc), whereas qualitative validation was done running
long simulations starting from the native structure of a 14-mer with a UUCG loop\cite{nozinovic2010high}.
\label{fig_systems}
}
\end{figure}

We here train the 12 gHBfix free parameters corresponding to the 12 types of hydrogen bonds (Fig.~\ref{fig_parameters}) by minimizing the discrepancy
with respect to the experiment for three systems:
two tetraloop motifs with sequence gcUUCGgc and gcGAGAgc and an oligomer with sequence GACC (Fig.~\ref{fig_systems}a).
The two tetraloops, or similar ones, were used as folding benchmark in a number of papers
(see, \emph{e.g.}, Refs.~\cite{chen2013high,kuhrova2013computer,bottaro2016free,kuhrova2016ghbfixfunc,Yang2017,tan2018rna,cesari2019fitting,kuhrova2019GHBFIX,mrazikova2020uucg,zerze2021}).
The GACC tetramer  was reported to sample intercalated structures not compatible with experiment \cite{Condon2015,bergonzo2015highly},
although this artifact can be significantly decreased using
modified dihedral potentials \cite{gil2016empirical,Aytenfisu2017} 
or modified water models \cite{bergonzo2015improved,Yang2017,bottaro2018conformational,tan2018rna}.
A hyperparameter that controls overfitting is tuned by mimimizing the cross-validation error.
Several different forms for the regularization term are compared.
Once an optimal value for the hyperparameter has been identified, a new fitting is performed including all the training
simulations, resulting in a set of optimal parameters that we refer to as gHBfix$_{opt}$.
We then test this set of parameters using additional simulations
that include new systems not used during training and a new simulation of one of the systems used during training (Fig.~\ref{fig_systems}b).
Furthermore (see Supporting Information), in order to identify side effects of the gHBfix$_{opt}$ parameters,
we perform plain MD simulations on carefully chosen folded RNAs (kink-turn and L1 stalk rRNA).
These additional simulations allow us to report a new
set of parameters (gHBfix21), where one hydrogen bond correction has been manually removed from the gHBfix$_{opt}$ parameters,
that performs well on a wider range of systems.

\subsection{Cross-validation comparison}

\begin{figure}
\includegraphics[width=0.45\textwidth]{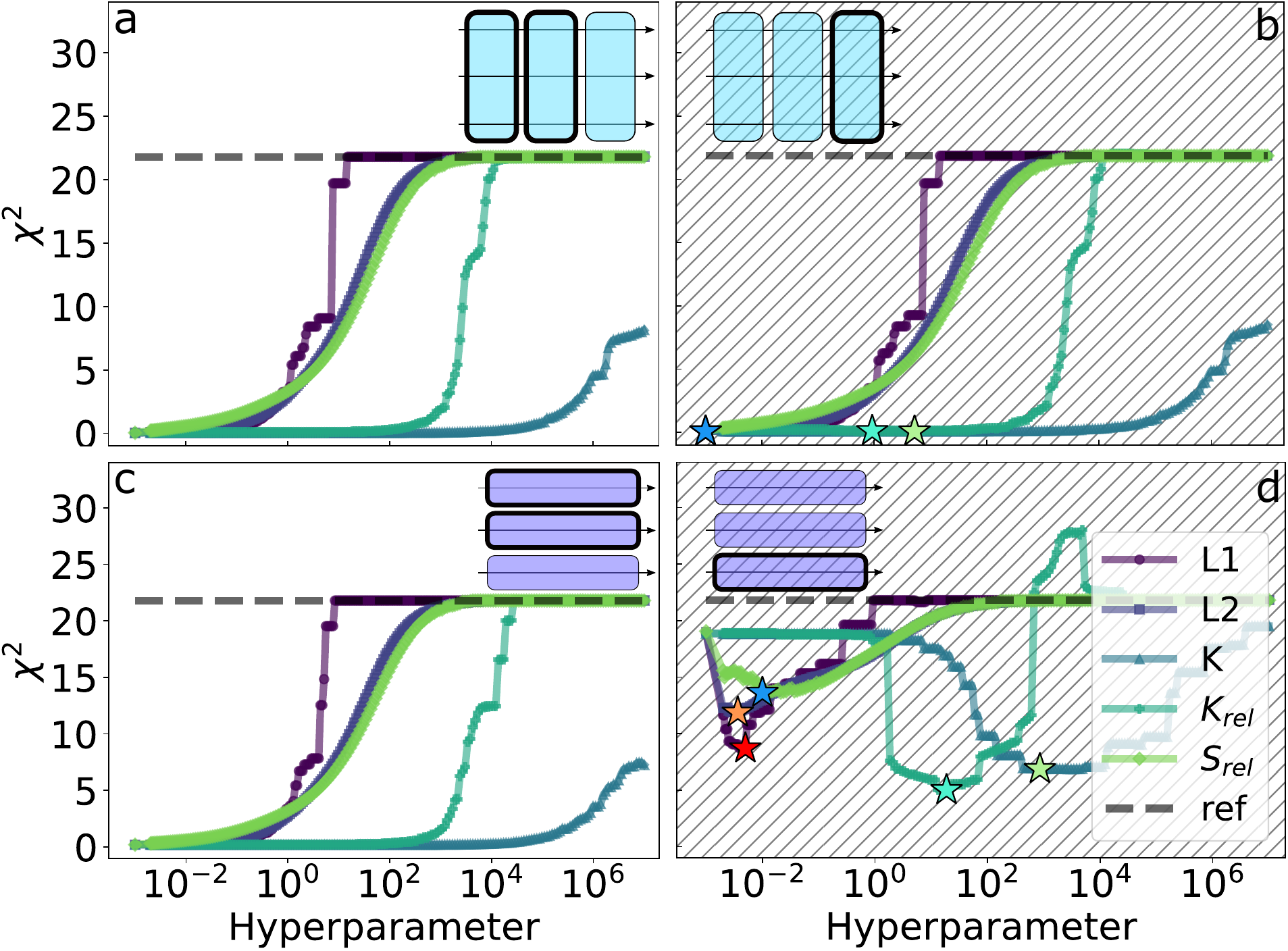}
\caption{
Results of the cross-validation tests on trajectory segments and on observables, using all the tested regularization methods.
The error function is evaluated on the training and validation set, using parameters obtained minimizing the error
and a scan over a wide range of values for the regularization hyperparameter.
Cyan and purple blocks schematize how data are split and used in cross validation.
In the first case (panels a and b), the cross-validation is performed keeping out segments of the whole trajectories and using them as a validation set.
Error is reported both for the training (panel a) and for the validation (panel b) set.
In the second case (panels c and d) the cross-validation is performed keeping out a fraction of the observables and using them as a validation set.
Error is reported both for the training (panel c) and for the validation (panel d) set.
Error function for the reference force field is reported as a dashed horizontal line.
\label{fig_results}}
\end{figure}

We first perform a cross-validation test on trajectory segments. In short, we split each trajectory in 5 segments, train the 12 parameters on a subset of 4 segments, and validate against the left-out segment. We repeat the procedure five times, and report the average result.
We then repeat the procedure scanning the value of the regularization hyperparameter over 8 orders of magnitude
and including 5 different forms for the regularization term.
Fig.~\ref{fig_results}a reports the average error on the training set.
By construction, the error increases with the hyperparameter.
For four forms of the regularization term, in the limit of large hyperparameters the error of the original force field
is recovered. When using the Kish size as a regularization term, instead, this is not guaranteed. Indeed, to maximise the
Kish size of the resulting ensemble, thus minimizing the regularization term, one should have uniform weights across
all the visited conformations, which is different from using the weights associated to the reference force field.
The limit of low hyperparameter corresponds to fitting without any regularization.
Fig.~\ref{fig_results}b reports the average error on the validation set, namely obtained using the trajectory segment
that was left-out during the training phase. In this case, the error systematically increases over a wide range of values of the
hyperparameter. The minimum error is not appreciably different from the error obtained
in absence of regularization. This indicates that, for what concerns the cross-validation on trajectory segments,
there is no significant overfitting, and we should expect the obtained parameters to be transferable to a new simulation performed on the same system irrespectively of regularization.

Fig.~\ref{fig_results}c and d instead report a similar analysis performed by splitting all the input datapoints in 7 groups corresponding to the observables,
training on a subset of 6 and validating on the left-out observable.
In this case, the behavior of the cross-validation error (Fig.~\ref{fig_results}d) is qualitatively different. In particular,
for each of the tested forms of the regularization term, we can clearly identify a specific value of the 
hyperparameter that minimizes the error on the validation set. For a low value of the hyperparameter instead,
we clearly see that the cross-validation error increases. This indicates that, in absence of regularization, one would obtain
parameters that would likely be non-transferable to predict new data points. 
The specific values of the hyperparameter that minimizes the cross-validation error are shown with a star.

\begin{figure}
\includegraphics[width=0.45\textwidth]{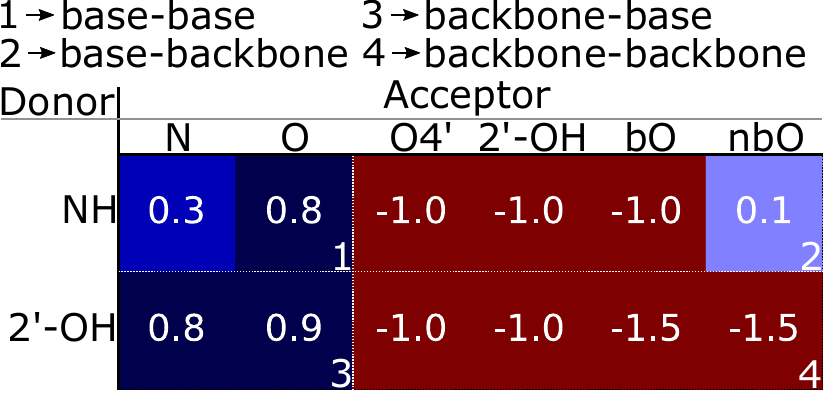}
\caption{
Final set of the 12 trained parameters (\emph{i.e.}, gHBfix$_{opt}$) that were obtained fitting on all the systems in the training set
using a relative Kish size regularization with a hyperparameter obtained by minimizing the error function of the cross-validation on observables.
Parameters are reported as $k_BT\cdot \lambda$ in kcal/mol (corresponding to the $\eta$ parameters in the gHBfix nomenclature \cite{kuhrova2019GHBFIX}) and boxes are colored in blue for attractive terms or in red for repulsive ones.}
\label{fig_GHBFIXoptimal}
\end{figure}

We notice that, given the different nature of the 5 tested regularization functions, the specific value of the
hyperparameter cannot be directly compared. We can however compare the corresponding values of the cross-validation error, that suggests that
the maximum transferability would be obtained using a relative Kish size regularization with a hyperparameter $\alpha=$18.68.
Using this criterion to choose the type of regularization function is legitimate and is
equivalent to considering the type of regularization function as an additional categorical
hyperparameter that is optimized using the same cross-validation procedure. 
For further tests we choose the parameters obtained with relative Kish size regularization at this optimal regularization strength, that are also reported in Fig.~\ref{fig_GHBFIXoptimal}.

\begin{figure*}
  \includegraphics[width=\textwidth,height=10cm]{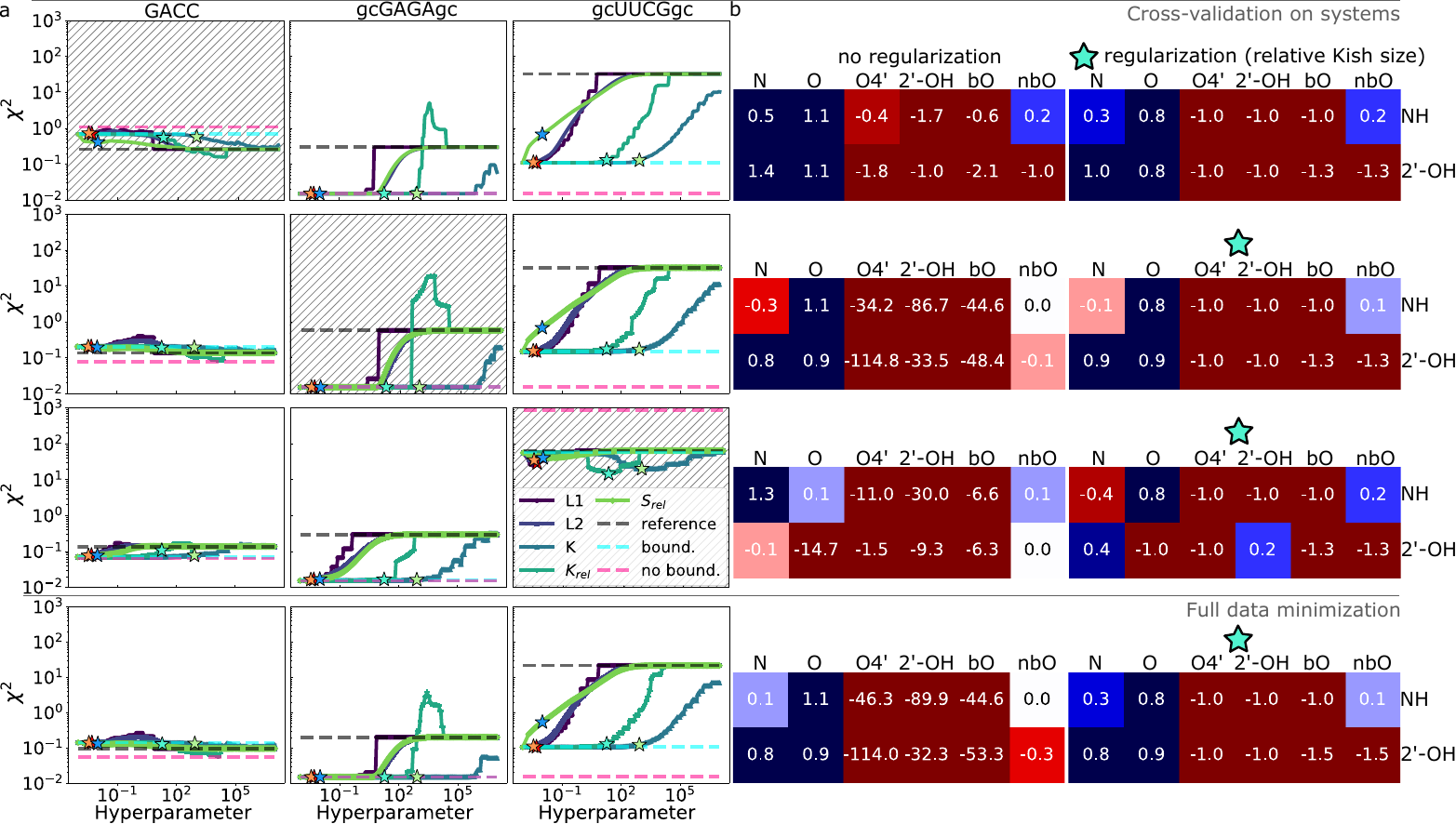}
  \caption{
Results of the cross-validation tests on multiple systems, using all the tested regularization methods (left panels).
The error function is evaluated on the training (white background) and validation (hatched) set, using parameters obtained minimizing the error
and a scan over a wide range of values for the regularization hyperparameter.
Error functions for the reference force field (gHBfix19), for an unregularized minimization without boundaries, and for an unregularized minimization with
boundaries at $\pm 1$ kcal/mol relative to the original force field are reported as horizontal dashed lines. 
Parameters associated to each minimization are shown in the right panels, using the same color scale as in Fig.~\ref{fig_overview},
both with and without regularization, as indicated. Parameters obtained with regularization are obtained using the hyperparameter indicated
with a star in the cross-validation plots and expressed as $k_BT\cdot \lambda$ in kcal/mol.
The last row shows the results for the training error using all available experimental and simulation data during the fitting (left)
and the corresponding parameters (right).
When regularized (bottom right panel), they correspond to the gHBfix$_{opt}$ parameters derived in this paper.
\label{fig_CVSys}}
\end{figure*}

We also performed a cross-validation on systems, where training is done including two systems and validation on the left-out
system (Fig.~\ref{fig_CVSys}).
This analysis allows to clearly identify the contribution of each of the three training systems to the resulting parameters.
The left panels show the error on each of the three analyzed systems, highlighting the one that was left-out during training.
The gcUUCGgc system has the highest error, as expected \cite{mrazikova2020uucg}.
Interestingly, gcGAGAgc is significantly improved by the presence of gcUUCGgc in the training set.
However, the opposite is not true: when gcUUCGgc is excluded from training, the associated validation error
displays a minimum that is almost as large as the error in the reference force field.
This suggests that the gcUUCGgc native structure is stabilized by types of contact not present in the other systems.
We notice that GACC tetramer shows a light overfitting whenever gcUUCGgc is included in the training set.
However, the magnitude of this overfitting is moderate, and the final $\chi^2$ error remains below 0.93.

It is also interesting to compare the sets of parameters obtained when regularization is present or absent,
and when one of the systems is left out or all the three systems are used (see Fig.~\ref{fig_results}).
When the GACC tetramer is included in the training phase and no regularization is used, some of the parameters become very large and negative.
For all considered systems these parameters correspond to hydrogen bonds that are more abundant in the fraction of the ensemble that is less compatible with the experimental observables, and thus should be penalized to improve the result. The particular importance of the concerted effect of these repulsive interactions on the correct representation of GACC can be seen in SI Fig.~1, which shows that removing all repulsive interaction to gHBfix19 parameter significantly increases the $\chi^2$ error of GACC to values $>1$.
Importantly, as soon as a regularization term is introduced, the obtained parameters are similar
irrespectively of which system has been left out from the training.
A fit performed using all the three systems with regularization also reports a similar result
(see Fig.~\ref{fig_results}).

The parameters obtained in all the tested minimizations are similar but not identical in the choice of which interactions should
be disfavored and which should be supported. In particular,
it emerges that NH-O base-base and 2$^\prime$-OH N/O sugar-base hydrogen bonds should be supported whenever gcUUCGgc is included
in training. The former type corresponds to Watson-Crick hydrogen bonds in the stem and GU wobble pair in loop, whereas the second type corresponds to signature interaction of the UUCG motif \cite{mrazikova2020uucg,westhof2005sigcont}.
In SI Fig.~2 one can see, that all attractive interactions
correspond to contacts that are present in the UUCG native loop.
Attractive interactions 2$^\prime$OH-N/O are exclusively present in the loop region,
thus are particularly helpful in correctly stabilizing the challenging UUCG motif.
In general, all the hydrogen bonds formed by acceptors located in the sugar or phosphate moieties should be disfavored,
with the notable exception of bonds between non-bridging oxygens and NH groups.
This might be a consequence of the limited set of systems analyzed in this work, where these moieties are
not involved in forming important interactions, as discussed in the next section.

\subsection{Tests using new simulations and additional systems}

The parameters obtained in the previous section were thus tested on new simulations.
We decided to use the parameters that were obtained fitting on all the training systems
using a relative Kish size regularization with a hyperparameter obtained by minimizing the error function of the cross-validation on observables.
These parameters are reported in Fig.~\ref{fig_GHBFIXoptimal} and are referred to as gHBfix$_{opt}$. As discussed in the previous section, however, these parameters are largely similar to other sets of parameters obtained in the cross-validation tests. 

The first test simulations were performed on three systems (Table \ref{tab_ValSys}). In particular, we tested two other tetramers (CAAU and UUUU), for which NMR data are available. CAAU is one of the most challenging tetramers \citep{Condon2015,bottaro2016rna,Aytenfisu2017,Yang2017,tan2018rna,bottaro2018conformational,kuhrova2019GHBFIX,Zhao2020,mlynsky2020fine}  and shows a relatively large $\chi^2$ of 5.37 with gHBfix19\cite{mlynsky2020fine} as well as 5.0 when reweighting the simulations using gHBfix$_{opt}$. The test simulation with gHBfix$_{opt}$ reduces the $\chi^2$ value to 3.6, which is an improvement. 
The UUUU system is relevant since it is known to be highly dynamic in NMR experiments \cite{Condon2015}
and was suggested to be too helical in Ref.~\cite{tan2018rna}.
Here the error after reweighting the simulations to gHBfix19 is at $1.7$, and the simulations with optimized parameters reported here lead to a decreased $\chi^2$ error as well.
These results are remarkable in that these two systems were not included in the training set.
In addition, we performed new simulations to validate our improvement on the UUCG tetraloop.
Specifically, a new ST-MetaD folding simulation of the gcUUCGgc system, using gHBfix$_{opt}$ corrections, but otherwise identical settings as the one used in the training phase,  resulted in a native state population of 21 $\pm$ 2\%. This number is comparable with the native state population of 27 $\pm$ 4\% estimated when reweighting the training simulation.
Whereas this system was used during training, it is interesting that the parameters were transferable to a direct (non reweighted) simulation,
consistently with what we observed in the cross-validation test on trajectory segments (see Fig.~\ref{fig_results}d).
Additionally, we performed a qualitative test on the stability of the UUCG tetraloop by performing
10 independent 20 $\mu$s plain MD simulations of a 14-mer initialized in an NMR structure \cite{nozinovic2010high}.
From SI Fig.~3 it can be seen, that in 9 out of 10 simulations, the native tetraloop structure
with all signature interactions \cite{mrazikova2020uucg} is stable.
We prolonged the single simulation where the native state was partly lost after $\approx$19 $\mu s$ and we observed
a successful recovery of all contacts at $\approx$20.4 $\mu s$. The native state was then maintained until the end of this 30 $\mu s$-long simulation.
This is in contrast with the instability observed with other variants of the AMBER force field
(see, \emph{e.g.}, Refs.~\cite{bottaro2016freecorrection,mrazikova2020uucg})
and with the local conformational dynamics observed with the parameters reported in Ref.~\cite{tan2018rna} (see
Refs.~\cite{kuhrova2019GHBFIXcorrection,bottaro2020integrating}).

In addition to these systems, we also choose two additional systems for which one specific type of hydrogen bonds that was penalized in
our fitted parameters was present in the native structure, namely 2$^\prime$OH-2$^\prime$OH bonds.
Results are reported in SI Fig.~5 and
strongly suggest that the disfavoring of the 2$^\prime$OH-2$^\prime$OH contact in gHBfix$_{opt}$ produces undesirable side effects.
We thus suggest that the 2$^\prime$OH-2$^\prime$OH term
should be set to 0.0 kcal/mol (\emph{i.e.}, removed) in simulations of any systems with A-minor, phosphate-in-pocket, and similar interactions with sugar-sugar H-bonding.
Our results on the kink-turn and L1 stalk rRNA segments confirm that, after removing this  2$^\prime$OH-2$^\prime$OH penalty, no side effects are observed on their native structures.
Importantly, the omission of the  2$^\prime$OH-2$^\prime$OH correction in gHBfix21 would not visibly compromise the results for the systems that
we have used in training or validation (see SI Fig.~4).
Remarkably, the test simulations also revealed that the stabilization of the 2$^\prime$OH-N H-bonds suggested by the fitting done on our training set further stabilizes the native kink-turn structure,
with respect to the uncorrected OL3 force field. Namely, the 2$^\prime$OH-N term eliminates dynamical bifurcation of the most important kink-turn signature interaction between the 2$^\prime$OH group of the first bulge nucleotide and N1 of the first adenine from the non-canonical stem \cite{Krepl2013}.

\begin{table*}
\centering
	\caption{
$\chi ^2$-errors and native populations for the training and testing simulations.
For the training simulations, we report both the direct results of the simulations (gHBfix19 column) and the results
predicted by reweighting those simulations to the gHBfix$_{opt}$ parameters displayed in Fig.~\ref{fig_GHBFIXoptimal}) (gHBfix$_{opt}$ column).
Notice that for GACC the results are obtained combining simulations performed with multiple parameter sets (see text for details).
For the validation simulations, we report both the direct results of the simulations (gHBfix$_{opt}$ column) and the results
predicted by reweighting those simulations to the gHBfix parameters with reweighting (gHBfix19 column).
}
\def\arraystretch{1.3}
\begin{tabular}{ c|c|c   }
\textbf{System (observable)} & \textbf{gHBfix19} & \textbf{gHBfix$_{opt}$} \\
\hline 
\multicolumn{3}{c}{\textit{Training simulation}} \\
\hline 
GACC ($\chi ^2$-NMR) & 0.24 \textsuperscript{a,b} & 0.33 \textsuperscript{d}\\
\hline
gcGAGAgc (native population)  &  24 $\pm$ 1\% \textsuperscript{c} & 66 $\pm$ 3\% \textsuperscript{d}\\
\hline
gcUUCGgc (native population) &  0.02 $\pm$ 0.002\% \textsuperscript{c} & 27 $\pm$ 4\% \textsuperscript{d}\\
\hline 
\multicolumn{3}{c}{\textit{Validation simulation}} \\
\hline
gcUUCGgc (native population) &  0.003 $\pm$ 0.004\% \textsuperscript{d} & 21 $\pm$ 2\% \textsuperscript{c,e}\\
\hline
CAAU ($\chi ^2$-NMR) & 5.00 \textsuperscript{d} (5.37 \textsuperscript{f}) & 3.64 \textsuperscript{b,e} \\
\hline
UUUU($\chi ^2$-NMR) & 1.73 \textsuperscript{d} (1.63 \textsuperscript{f})& 1.48 \textsuperscript{b,e}  \\
\end{tabular}
\begin{minipage}[t]{0.5\linewidth}
\raggedright
\textsuperscript{a} Results are obtained combining simulations performed with multiple parameter sets (see methods for details). \par
\textsuperscript{b} REST2 simulation.\par
\end{minipage}\hfill
\begin{minipage}[t]{0.5\linewidth}
\raggedright
\textsuperscript{c} ST-metadynamics simulation.\par
\textsuperscript{d} Reweighted results.\par
\textsuperscript{e} Simulations with gHBfix$_{opt}$ parameters.\par
\textsuperscript{f} Simulations with gHBfix19 parameters \cite{mlynsky2020fine}. %
\end{minipage}
\label{tab_ValSys}
\end{table*}

\section{Discussion}

In this work, we apply a force field fitting strategy that was introduced in a previous work \cite{cesari2019fitting}
to the tuning of gHBfix hydrogen bond interaction terms that were introduced in Ref.~\cite{kuhrova2019GHBFIX},
obtaining parameters that we call here gHBfix$_{opt}$.
Specifically, experimental data for two tetraloops and one tetramer are used to fit corrections that are then
tested on newer simulations of one of the two tetraloops and on two tetramers not seen during fitting.
The obtained parameters result in a significant stabilization of the difficult UUCG tetraloop.
Additional tests are performed on systems where 2$^\prime$OH-2$^\prime$OH H-bonds are essential for stabilization, since this H-bond is not present in any of the RNA structures considered in the training set, indicating general applicability of the gHBfix21 parameter set on a wide range of RNA systems. Using gHBfix21 instead of  gHBfix$_{opt}$ does not compromise the performance for any of the training and validation systems while it eliminates all side-effects on the native structures of kink-turn and L1 stalk rRNA.
Scripts that can be used to repeat the fits and reproduce the figures of this article can be found at \url{https://github.com/bussilab/ghbfix-training}.

When compared with Ref.~\cite{cesari2019fitting}, we report a number of methodological improvements.
First, we test five different regularization strategies. Two of them (L1 and L2) are standard in the machine learning
community and can be directly interpreted as prior distributions on the parameters aimed at keeping them small (L2) or sparse (L1).
We also test two additional strategies that are aimed at keeping the resulting reweighted ensemble as close as possible
to the original one (relative entropy and relative Kish size). Interestingly, there is an analogy between using the relative entropy
as a regularization term and the Bayesian experimental restraints introduced in Ref.~\cite{hummer2015bayesian}. Indeed, in both cases, among the multiple possible ensembles that are equally in agreement with experiment, the method will pick the one that is as close as possible to the original ensemble.
At variance with Ref.~\cite{hummer2015bayesian}, however, the approach introduced here is aimed
at deriving transferable corrections.
Finally, we test the possibility to regularize using the inverse of the Kish size, which allows to keep the resulting
reweighted ensemble as statistically rich as possible. A similar idea was proposed in Ref.~\cite{kofinger2021empirical}, though
using a different functional form. We notice that
in some cases the initial trajectories are generated using algorithms that provide conformations
associated with a weight. This happens, for instance, when using enhanced sampling methods where a bias is applied
or when combining trajectories obtained with different force fields using binless weighted histograms
\cite{Souaille2001,Shirts2008,Tan2012}.
In these cases,
using a Kish size regularization
makes the ensembles as uniform as possible, and thus the result might depend significantly on which
ensembles were sampled originally and which enhanced sampling strategy was used.
In the last three discussed strategies (relative entropy, relative Kish size, and Kish size), the penalty introduced by the regularization
term does not depend only on the parameters but also on the data,
and thus can be interpreted as a form of representational regularizations \cite{goodfellow2016deep,calonaci2020machine}.

Using as fitting coefficients the prefactors associated to the gHBfix corrections \cite{kuhrova2019GHBFIX},
makes their interpretation straightforward, as they directly report on how much each hydrogen bond type is to be
supported (positive coefficient) or penalized (negative coefficient).
When one of the trained coefficients diverges, the weight of frames where one interaction of that type is either present (for a negative coefficient)
or absent (for a positive coefficient) are effectively removed from the ensemble. The result is thus mildly depending on the
exact value of the coefficients. This means that, for selected training sets, one or more of the parameters might diverge
with some of the regularization strategies mentioned above.
This might lead to forces of infinite magnitude if these corrections were applied to a new simulation.
To avoid this type of issues, we added
a L-infinite-like regularization term that enforce all the parameters to be within preassigned boundaries. Namely, we favor or disfavor any of the corrected pairs by at most 1 kcal/mol.
The possibility to automatically repeat the training using different subsets of systems allows one to judge the contribution of each system
in the overall fitting.
Similarly, it is easy to repeat the training manually removing some of the correction, so as to identify the role of each term.

Another concept that is introduced here is that of performing a cross-validation over trajectory segments. This allows one
to assess how much the parameters would be generalizable to a new trajectory for the same systems.
This is a useful criterion to decrease the impact of errors due to finite sampling.
In our dataset, even in absence of regularization, no significant overfitting on the trajectory segments emerges,
indicating that our trajectories are long enough to be used in this training procedure. However,
for more complex systems or for shorter trajectories, this might not be true.

The optimized parameters, that we refer to as gHBfix$_{opt}$ parameters, perform well both in a new simulation of the difficult UUCG tetraloop
and in the simulation of two tetramers not seen during training, confirming that parameters are transferable.
For the UUCG tetraloop, we remark that the native state population reported here is higher than the one reported in Ref.~\cite{tan2018rna}.
Importantly, the G${_{\text{L4}}}$ bulge out structure that has been described both for the parameters
of Ref.~\cite{tan2018rna} (see Refs.~\cite{bottaro2019barnaba,bottaro2020integrating,kuhrova2019GHBFIXcorrection})
and for previous variants of the AMBER force field \cite{kuhrova2013computer,bergonzo2015highly,cesari2019fitting},
and that is not compatible with experimental solution data \cite{nozinovic2010high,nichols2018high,bottaro2020integrating},
is not populated in our plain MD simulations.
The capability of the flexible functional form of the gHBfix correction to directly stabilize the signature interactions
present in the native structures with no or minimal side effects,
coupled with the explicit inclusion of a UUCG tetraloop in our training set,
allows for the required corrections to be automatically detected.
We speculate that this result can be only
achieved with such a flexible functional form.
A folding simulation using the proposed parameters and the full 14-mer system
is left as a subject for a future work.

It is additionally important to notice that some interaction types were not present in the native structures
of the systems used in our training set. These interactions were thus maximally penalized by the training procedure.
Particularly relevant is the case of interactions between a pair of 2$^\prime$OH groups.
Sugar-sugar H-bonding is an important component of A-minor and all other types of ribose zipper interactions
\cite{sponer2018rna,cate1996crystal,batey1999tertiary,tamura2002sequence,mokdad2006structural}.
Sugar-sugar interactions are omnipresent in folded RNAs and the A-minor interaction is actually the most abundant RNA tertiary interaction used by evolution \cite{sponer2018rna,nissen2001rna}.
These interactions are indeed crucial for maintaining, for example, the native fold for a kink-turn motif and for the L1 stalk rRNA, which in turn includes
two kink-turn motifs.
In order to simulate systems where these interactions play an important role, the optimized parameters should be manually modified.
In theory, one could directly include kink-turns in the training set. However, trajectories where the native structure
is folded and unfolded at equilibrium would be required to estimate the effect of a correction on the stability of the
native structure using a reweighting procedure. Whereas this might be possible at least for the kink turn studied here,
it would be extremely expensive and will be left as a subject for a future work. We here decided to manually remove a single parameter,
and to validate it on the kink-turn motif using standard MD simulations. The resulting gHBfix21 parameter set is proposed to be applicable on a wider range of systems.

We notice that with the present gHBfix$_{opt}$ version the kink-turn and L1 stalk rRNA were significantly destabilized, however not to the same extent as with another
recent reparametrization of the RNA force field \cite{tan2018rna}, as shown in Ref.~\cite{kuhrova2019GHBFIX}.
This might be related to the fact that the parameters introduced in Ref.~\cite{tan2018rna} were
optimized to correctly fold structural motifs similar to the tetraloops used in our training set.
This observation corroborates the fact that the training set should be as heterogeneous as possible to avoid overfitting \cite{frohlking2020toward}.
However, the flexibility of the method allows parameters to be adjusted so as to manually remove some of the terms and train again
the remaining ones. With the adjusted gHBfix21 parameter set all side-effects on the kink-turn and L1 stalk rRNA were eliminated.

An important advantage of the gHBfix functional form is its modularity, namely the fact that it is possible to act on specific hydrogen bond types while
minimizing the indirect effects on others.
In fact, it is separated from all the other force-field terms. 
In order to allow flexible adjustments to the gHBfix$_{opt}$ parameter based on some detailed system knowledge, in the SI we show fitted force-fields which are omitting specific interactions, or reduce upper and lower bound of the parameters during fitting. Additionally, the SI provides a fitting script, which allows users to specify interactions to remain unchanged or within a certain magnitude and find a new force field matching these requirements (SI 8.5).
In case one is concerned about too large changes of the relative stability of AU and GC pairs with the gHBfix$_{opt}$ parameters, 
in SI Fig.~4 we offer a gHBfix$_{opt}$-version with a reduced magnitude of the NH-O interaction and we also show its expected effects on the training set.
In other words, the users can modify the gHBfix in specific projects in a system-specific manner.
We recall that the present gHBfix21 version was derived to be applied on top of the basic AMBER OL3 RNA force field
\cite{Wang2000,Perez2007,Zgarbova2011},
with phosphate oxygen corrections
\cite{Steinbrecher2012,Mlynsky2015} and
combined with the OPC water model \cite{IzadiOPC2014}.

Future studies should investigate whether non-linear functions can additionally improve force fields by allowing more functional flexibility, \emph{e.g.}, in the form artificial neural networks, when one attempts to find correction potentials for  more extensive databases of RNA dynamics.

\section{Acknowledgement}

This work was supported by Czech Science Foundation (20-16554S to V.M., M.K. and J.S.).

\begin{suppinfo}
Eight sections discussing 
(1) all simulations analyzed in this paper, 
(2) sensitivity analysis on gHBfix$_{opt}$, 
(3) favored interaction types of gHBfix$_{opt}$, 
(4) standard simulations of the 14-mer ggcacUUCGgugcc system carried out with
gHBfix19 and gHBfix$_{opt}$, 
(5) custom constraints on corrections in order to take into account possible side-effects, 
(6) MD simulations of kink-turns Kt-7 and L1 stalk rRNA,
(7) helical properties in simulations on RNA duplexes corrected with gHBfix$_{opt}$,
(8) the provided scripts allowing to obtain all published results of this study. 
\end{suppinfo}

\bibliography{main}%

\end{document}